\documentclass[twocolumn,showpacs,amsmath,amssymb]{revtex4}
\usepackage{latexsym}
\usepackage{graphicx}
\usepackage{pifont}
\usepackage{color}

\newcommand{\req}[1]{Eq.\,(\ref{#1})}

\newcommand{\beqn}{\begin{equation}}
\newcommand{\eeqn}{\end{equation}}

\newcommand{\Tmm}{T^{K}_{K}}
\def\lambe{\ensuremath{{\mathchar'26\mkern-9mu \lambda_e}}}

\begin{document}
\title{Spectra of Particles from Laser-Induced Vacuum Decay}
\author{Lance Labun and Johann Rafelski}
\affiliation{Department of Physics, University of Arizona, Tucson, Arizona 85721, USA}
\vskip 4mm

\begin{abstract} $\phantom{v}$

The spectrum of electrons and positrons originating from vacuum decay
occurring in the collision of two noncollinear laser pulses is obtained.
It displays high energy, highly collimated particle bunches traveling in a
direction separate from the laser beams.  This result provides an
unmistakable signature of the vacuum decay phenomenon and could suggest a
new avenue for  development of high energy electron and/or positron beams.
\end{abstract}

\maketitle

\section{Introduction}
Quantum electrodynamics implies that, when the potential difference in an electromagnetic field exceeds $2mc^2$ in constant~\cite{Heisenberg:1935qt,Schwinger:1951nm} and nonconstant fields~\cite{Sauter1932}, the  field-filled domain decays by particle emission.  Many other effects of the quantum-induced nonlinearity have been studied over the years, including photon-splitting~\cite{Adler:1970gg,BialynickaBirula:1970vy}, the energy-momentum trace~\cite{Adler:1976zt,Dittrich:1988yi,Labun:2008qq} and vacuum birefringence~\cite{BialynickaBirula:1981,Adler:2006zs}.  None of these effects has yet been experimentally observed.

A plane electromagnetic wave of arbitrary intensity and wavelength composition is absolutely stable against pair materialization decay~\cite{Schwinger:1951nm}, since there is no reference frame in which massive decay products can appear.  
Colliding laser pulses can form a field configuration that does have a preferred reference frame: the frame in which the local field momentum density (the Poynting vector) vanishes defines the local rest frame in which decay products can materialize.  The field energy present in this frame is the amount available for materialization~\cite{Labun:2008re}.  

In fact, any electromagnetic field configuration having an identifiable reference frame can decay, and will decay for any positive value of the invariant 
\beqn
d^2 = \sqrt{{\cal S}^2+{\cal P}^2}-{\cal S} > 0, 
\eeqn\vspace*{-6mm}\beqn
{\cal S} = \frac{1}{2}(\vec B^2-\vec E^2), \quad {\cal P} = \vec E \cdot \vec B
\eeqn
A previously in-depth studied example is pair production by a supercritical Coulomb potential near a $Z>173$ nucleus~\cite{Greiner:1985ce}.  There, the bare charge of the at-rest nucleus is screened by the charged vacuum state~\cite{Rafelski:1974rh} localized around the nucleus, and a positron escapes. If and when the supercritical potential dissociates, the localized electron charge from the charged vacuum state could also be freed, completing the pair formation process.

Many studies have addressed pair creation in intense laser pulses colliding head-on~\cite{Brezin:1970xf,Narozhny:1973af,Alkofer:2001ik,Fried:2001ga,Avetissian:2002,DiPiazza04,Blaschke:2005hs,Bell:2008zzb,Ruf:2009zz,Kirk:2009vk,Dunne:2009gi,Bulanov:2010ei,Dumlu:2010ua,Bulanov:2010gb} (since this offers the highest particle yield) or traveling parallel~\cite{Ringwald:2001ib,Bulanov:2004de,Schutzhold:2008pz,Kingetal,Dumlu:2010vv} (since chirped laser pulses consist of many focused frequency components). In this work we show how one obtains spectra of produced particles in the laboratory by identifying the frame of reference in which production occurs. We evaluate for the case of noncollinear colliding lasers, a field configuration whose local rest frame moves at high rapidity with respect to the laboratory. The advantages of this geometry are that
(1) nearly monochromatic bunches of high energy particles are produced, and 
(2) the bunches are boosted in the direction of the Poynting vector of the combined laser pulses and thus away from each individual laser pulse.  
These characteristics constitute a signal that the pairs originate in vacuum decay, which from the standpoint of discovery potential outweighs the reduction in particle yield as compared to head-on collisions.
Mirroring the practice in elementary particle physics, the decay rates and products are studied in this work in terms of Lorentz invariants, keeping separate the particle production process from the reference frame determination.


\section{Rapidity of Moving Rest Frame}
In the laboratory frame characterized by a 4-velocity $u^K=(1,\vec 0)$ the 4-momentum density is $p^K=T^{KJ}u_{\!J}$ using the energy-momentum tensor $T^{KJ}$. The energy density $p^0$ and the Poynting momentum density $p^i$ of an electromagnetic field configuration are
\beqn\label{4momentum}
p^0=\varepsilon\frac{\vec E^2+\vec B^2}{2}+\frac{\Tmm}{4}, \qquad 
p^i = \varepsilon(\vec E\times \vec B)^i =: \vec S.
\eeqn
The dielectric shift $\varepsilon-1$ and the energy-momentum trace $\Tmm$ vanish in classical electromagnetism and acquire nonzero field dependent values in QED due to fluctuations in the vacuum~\cite{Labun:2008qq}.

The (square) invariant mass density $\mu^2=u_{\!J}T_{K}^{J}T^{KJ'}\!\!u_{\!J'}$ can be studied in any frame of reference, in particular in the lab frame or in the frame having $\vec S = 0$ where energy density is rest mass density.  We proceed to evaluate mass density in the lab frame.  Using the explicit expression for the electromagnetic energy-momentum tensor, we obtain~\cite{Labun:2008qq,Labun:2008re}
\beqn\label{muEM}
\mu^2:= p_K p^K = \varepsilon^2({\cal S}^2+{\cal P}^2) + (\Tmm/4)^2.
\eeqn

For general field configurations, $\mu$ is a function of space and time and determines the instantaneous, local rapidity $y_{\rm S}$ of the moving rest frame 
\beqn\label{eta}
\cosh^2y_{\rm S} = \frac{\mu^2+\varepsilon^2(\vec E\times \vec B)^2}{\mu^2}
=1+\frac{\vec S^2}{\mu^2}
=\frac{(p^0)^2}{\mu^2} 
\eeqn 
which exhibits the dependence on the geometric relation between the electric and magnetic field vectors.  As for particles, large electromagnetic momentum density $|\vec S| > \mu$ in a given frame corresponds to high rapidity. Note that the choice $u^K=(\cosh y_{\rm S}, 0, 0,\sinh y_{\rm S})$ produces the corresponding result for mass density working in the local center of momentum frame. 

An isolated electric field is at rest in the laboratory.  In the presence of any magnetic field along with the electric field, the total field configuration acquires momentum dependent on the angle between the electric and magnetic field vectors, seen in the nonzero Poynting vector \req{4momentum}. If the magnetic field is orthogonal to the electric field (${\cal P}=0$), a frame exists where one of $\vec E,\vec B$ vanishes and is the rest frame.  Otherwise, when ${\cal P}\neq 0$ the frame where $\vec E,\vec B$ are parallel is the rest frame.

Consider the example of a linearly polarized plane wave entering a domain of quasiconstant magnetic field.  The largest quasistatic magnetic fields in laboratory are small in comparison to the fields available in pulsed laser systems, and the ratio of the external field to the laser field strength
\beqn
\frac{eB_{\rm ext}}{a_0\omega m} <10^{-4}, \quad  
a_0 := \frac{e|A^{K}|}{m}=\frac{e|\vec E_{\rm laser}|}{\omega m}
\eeqn
where $a_0$ is the dimensionless laser amplitude and $\omega$ is the laser frequency.  The resulting superposition of fields has a small mass density and a large  rapidity.  The same choices for field geometry that provide a large $y_{\rm S}$ reduce the magnitude of $\mu$ due to the finite total energy density being apportioned between rest mass density and momentum density by $\cosh y_{\rm S}$. 

For a present day high-intensity laser with $\omega \simeq 1\:{\rm eV}$ and $a_0=100$ and an external magnetic field of $1\:{\rm T}$ aligned with the laser polarization, the resultant mass density $\mu(x,t)$ and rapidity $y(x,t)$ at each position and time are
\beqn
\mu(x,t) \lesssim 5\:10^8\:\frac{\rm MeV}{\rm nm^3}, \quad y(x,t) \gtrsim 14.
\eeqn
This $\mu$ is 14 orders of magnitude smaller than the scale set by the electron  $0.511~{\rm MeV}/\lambe^3= (e\vec{E}_c)^2$ where $|\vec E_c|={m^2c^3}/{e\hbar}  = 1.3\times 10^{18}\:{\rm V/m} = 4.4\times 10^9c\:{\rm T}$, and too small to generate a significant number of electron-positron pairs.

\section{Materialization of Electron-Positron Pairs}
Decay of the unstable field into an electron-positron pair is treated as a semiclassical tunneling process~\cite{Casher:1978wy}, which shows the momentum distribution to be dominantly Gaussian.  Because of the adiabatic $\omega\ll m_e$ turn-on of the field, particles are created at threshold with zero momentum $p_\parallel=0$ in the direction of the electric field in the field's moving rest frame, and the energy $E_p$ of the outgoing particle depends only on the momentum $\vec{p}_\perp^{\,2}$ perpendicular to electric field.
The tunneling probability
\beqn\label{tunnel}
\Gamma \propto  e^{-\pi E_p^2/ed}, \qquad E_p^2 = \vec{p}_\perp^{\,2}+m^2.
\eeqn
and other pertinent quantities  depend  on the field invariants
$ d^2\! =  \sqrt{{\cal S}^2+{\cal P}^2}\! - {\cal S} \to |\vec E|^2 $ and
$ b^2 \!=  \sqrt{{\cal S}^2+{\cal P}^2} \!+ {\cal S} \to |\vec B|^2 $ 
which reduce to the corresponding electric and magnetic magnitudes in the moving rest frame of the field. 
Deriving the spectrum requires knowing the orientation of the field: $\vec{p}_{\perp}$ above is the electron (positron) momentum  transverse to the direction of the electric field vector present in the moving rest frame of the decaying field. Note that transforming to the laboratory frame preserves the Gaussian shape. 

The total number of expected pairs $\langle N\rangle$ is obtained by integrating \req{tunnel} over transverse momenta and multiplying by total volume and time over which the unstable field configuration ($E^2-B^2>0$) exists~\cite{Nikishov:1970br,Cohen:2008wz} 
\begin{align}
\frac{1}{VT}\langle N\rangle &= 
\frac{e^2bd}{8\pi^2}\coth\!\left(\!\frac{\pi b}{d}\!\right)e^{-\pi m^2/ed} 
\underset{b\to 0}{\longrightarrow} \frac{(ed)^2}{8\pi^3}e^{-\pi m^2/ed},
\end{align}
The rate at which field energy is converted into material particles is computed by weighting the momentum integral with the energy of the particle, giving~\cite{Labun:2008re} 
\beqn\label{dumdt}
\frac{d\langle u_m\rangle}{dt} 
 \simeq  \frac{\alpha m}{\pi^2}d^2e^{-\pi m^2/ed}.  
\eeqn
The characteristic time for conversion of field energy into particles usually far exceeds the laboratory duration of the field.  For $d\gtrsim 0.1|\vec{E}_c|$ however, pair creation becomes copious and the lifetime approaches the  time scale of the laser pulses composing the field.   Consequently, the fraction of field energy converted into particles becomes large and detailed treatment of backreaction is necessary~\cite{Cooper:1992hw,Mihaila:2009ge}.

\section{Colliding Laser Pulses and Particle Spectra}
\begin{figure}
  \includegraphics[width=0.46\textwidth]{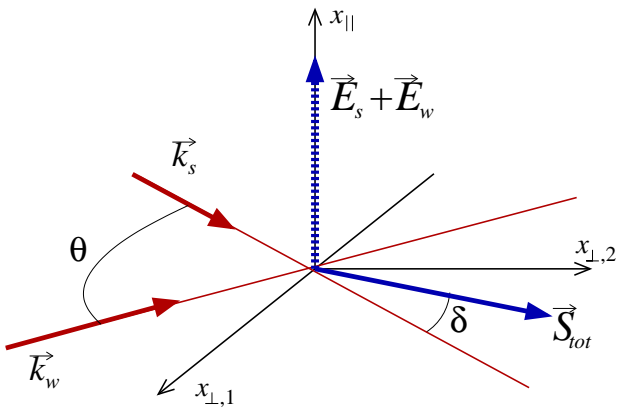}
\caption{(color online)  Schematic of two converging linearly-polarized laser waves, $\vec k_{\rm s},\vec k_{\rm w}$, with $\vec E_{\rm s},\vec E_{\rm w}$ the corresponding electric field vectors.  $\vec S_{\rm tot}$ is the Poynting vector of the resultant field, deflected an angle $\delta$ by the second laser pulse.}
\label{fig:vectors}
\end{figure}

As the strongest available electromagnetic fields are today created with pulsed lasers, colliding pulses offers the best means to maximize $d$ and create unstable field configurations.  Judicious choice of collision geometry creates a region where both the rapidity $y$ and the electric invariant $d$ are large.  Figure \ref{fig:vectors} illustrates two linearly-polarized laser waves $\vec k_{\rm w},\vec k_{\rm s}$  (subscript w for ``weak'' and s for ``strong''), converging at an angle $\theta$ with electric field vectors $\vec E_{\rm s},\vec E_{\rm w}$ aligned in the $x_{\Vert}$-direction.  The wave vectors $k_{\rm w},k_{\rm s}$ define the $x_{\perp,1}$-$x_{\perp,2}$ collision plane, in which the collision angle $\theta$ varies between exactly collinear laser pulses $\theta=0$ and a head-on collision $\theta=\pi$.

The Poynting momentum of the total field $\vec S_{\rm tot}$ is determined by the net magnetic field: the vector sum of the lasers' magnetic fields deflects $\vec S_{\rm tot}$ away from the beam line of the strong laser by an angle $\delta$ given by 
\beqn
\tan\delta = \frac{r\sin \theta}{1+r\cos\theta}; 
\quad 
r = \frac{a_0^{\rm w}\omega^{\rm w}}{a_0^{\rm s}\omega^{\rm s}}\le 1
\eeqn
$\vec S_{\rm tot}$ is in the plane normal to the electric field (the collision plane) also when considered in the moving rest frame of the field.
$\vec{p}_{\perp}$ of produced particles is in this collision plane. Upon boosting by rapidity $y_{\rm S}$ to the lab frame, the Gaussian distribution, which is uniform in $\vec{p}_{\perp}$ in the moving rest frame, is shifted in one of the two transverse components, remains centered around zero in the other transverse momentum component and remains negligible in direction parallel to the electrical field. 

For a qualitative model, we take square-wave-localized pulses.  Setting the waves to be in phase, the oscillation within each individual pulse results in a checker-board of stable ($E^2-B^2<0$) and unstable ($E^2-B^2>0$) subdomains in the volume where the pulses overlap.  The stable subdomains do not contribute to the particle output.  Within the unstable domains the individual sinusoidal variations of the laser fields produce a variation in the direction and magnitude of $\vec S_{\rm tot}$.  However, most particles are created near the peak of the sinusoid where variation of $\vec S_{\rm tot}$ is slow.  Space and time dependence of the fields affects particle yields by a factor of order unity and spectra by small changes to distribution widths.  In the following, we evaluate using the peak field magnitudes and geometry.

\begin{figure}  
  \includegraphics[width=0.48\textwidth]{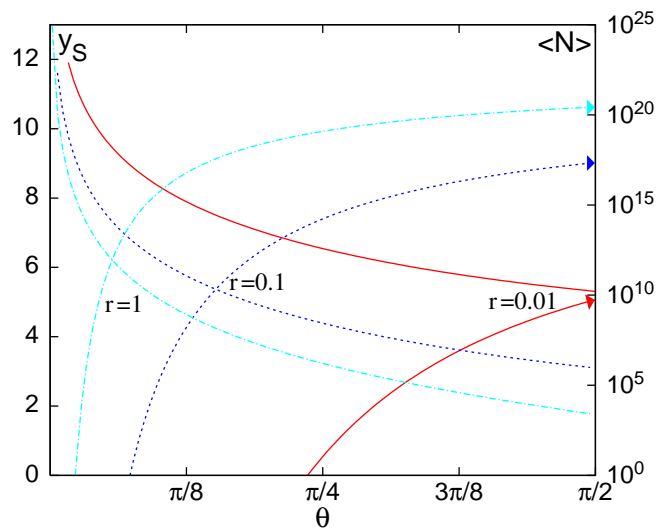}
\caption{(color online)  The rapidity $y_{\rm S}$ of the moving rest frame (left scale) and expected number of pairs (right scale, lines with arrows) for $r=1, \ 0.1,\ 0.01$ at fixed $a_0^{\rm s}\omega^{\rm s}/m=1$. }
\label{fig:eta}
\end{figure}

In unstable subdomains, the magnetic field is reduced by $\cos\theta$ in magnitude compared to the net electric field and remains orthogonal to it.  As a result, 
\beqn
b^2=0,\quad (ed)^2=2r(1-\cos\theta)(a_0^{\rm s}\omega^{\rm s} m)^2.  
\eeqn
We consider three cases: equal intensity $r=1$, and `weak deflections' $r=0.1,\ =0.01$. 
The rapidity $y$ is shown on the left axis in Fig.~\ref{fig:eta} with corresponding three lines merging at small $\theta$ at large $y$ (from top to bottom: small $r$ to $r=1$).  For the collision of equal intensity pulses $r=1$ the rapidity goes to zero approximately linearly as $\theta\to\pi$, reflecting that standing waves created by counter-propagating square laser pulses are at rest in the laboratory.  In the asymmetric $r<1$ cases, the weak laser field serves similar to the quasiconstant external field example above, giving the strong laser pulse a mass density and the ability to decay at relatively high rapidity $y_{\rm S}$.

To create pairs with initially high rapidity in the lab, we focus on $\theta<\pi/2$.  The number of pairs obtained would be maximized for the head-on collision $\theta=\pi$, but in this case the pair-producing field is at rest in the lab and the pairs appear with zero net momentum.  Decreasing $\theta$ results in higher rapidities, but in order to maintain the same yield in particles, intensity of the field in the collision domain must be increased, whether by increasing the intensity of the strong pulse $a_0^{\rm s}$ or the ratio $r$.  In the limit $\theta\to 0$, the mass density (and hence the potential for pair production) vanishes because the converging laser pulses have degenerated into collinearly propagating waves.

As for the above example of a laser field incident on an external magnetic field, the trade-off between yield of produced pairs and rapidity of the moving rest frame is a consequence of the finite total energy density of the laser pulses.  This trade-off is visible in the $\theta$ dependence of the invariant mass density, which derives from nonvanishing $d^2$ in the region of superposition, 
\beqn\label{mu-tot}
\mu^2 = 2\varepsilon^2r^2(1-\cos\theta)^2+(\Tmm/4)^2.
\eeqn
The rapidity of the moving rest frame of the field is 
\beqn\label{gamma-coll}
\sinh^2\! y_{\rm S}\! = 
\frac{\left[(\frac{1}{2}(r^{-1}\!+r)\csc^2(\theta/2)+\cot^2(\theta/2)\right]^2\!
-1}{1+\left(\Tmm/4\varepsilon d^2\right)^2}
\eeqn
For the strong fields necessary to induce significant pair creation, $\Tmm/4$ is a noticeable correction, comparable in magnitude to focusing corrections.  In generating Fig.~\ref{fig:eta}, we have used nonperturbatively computed values of $\varepsilon$ and $\Tmm$~\cite{Labun:2008qq}.  For $d\lesssim 0.1|\vec E_c|$ their series expansions, given in~\cite{Labun:2008qq}, can be used to good accuracy.

\begin{figure}  
  \includegraphics[width=0.48\textwidth]{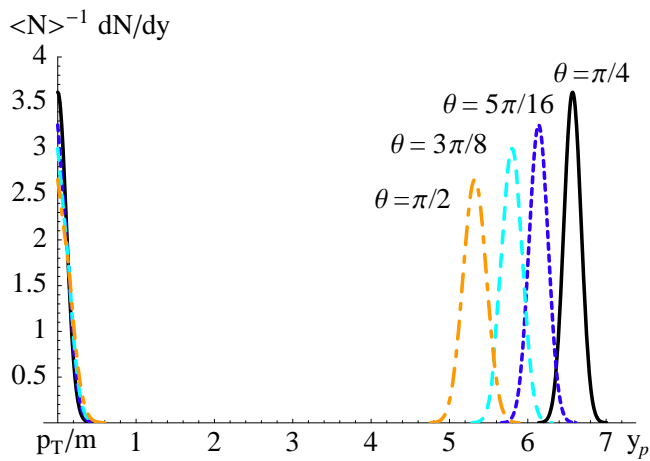}
\caption{(color online)  The normalized distributions in rapidity and transverse momentum for $a^{\rm s}_0\omega^{\rm s}/m = 1$, $r=10^{-2}$ and $\theta=\pi/2,\ 3\pi/8,\ 5\pi/16,$ and $\pi/4$.  To the left $p_{\rm T}/m$ spectra and to the right the rapidity spectra. See text for specific meaning of $p_{\rm T}$. }
\label{fig:spectrum}
\end{figure}

In the laboratory frame, the vacuum-derived pairs have a boosted mean momentum $\langle p_{\parallel{\rm S}}\rangle = \sqrt{m^2+p_{\rm T}^2}\sinh (y-y_{\rm S})$ in the direction of $\vec S_{\rm tot}$, thus defining  direction of the bunch.  In the plane normal to $\vec S_{\rm tot}$, the produced particles have a momentum distribution concentrated in the direction that is perpendicular also to the net electric field vector.  In this direction, the $p_{\rm T}$ momentum distribution is Gaussian with mean zero, unaffected by the Lorentz transformation along $\vec S_{\rm tot}$.  With high rapidity and low transverse momenta, the vacuum-decay products form a pancaked, tightly collimated bunch with low transverse emittance. Distributions in rapidity $y_p=y-y_{\rm S}$ and transverse momentum $p_{\rm T}$ are shown in Fig.~\ref{fig:spectrum} for four selected collision angles $\theta=8\pi/16, \ 6\pi/16, \ 5\pi/16,\ 4\pi/16$. To the left, we see the transverse momentum distribution. To the right, we see normalized distribution in $y_p$, integrated over $p_{\rm T}$.

The spatial size of the produced particle bunch is determined by the size of collision volume and hence for lasers focused near to the diffraction limit is $\sim\!\lambda^3 = (1\:{\rm \mu m})^3$.  Note that charge separation due to the definite orientation of the electric field implies that the positrons appear  separated from the electrons by about the width of the pulse.  As $\langle N\rangle$ is an extensive quantity, we compute reference values for a collision volume $\times$ duration of $(5\lambda)^3\times 25\lambda/c$.  Figure~\ref{fig:eta} exhibits the trade-off between $y,\langle N\rangle$ (right scale, lines with arrows), which arises from the trade-off between $y_{\rm S},\mu$ noted above. 

\section{Conclusions}
The field rest frame is general to classical electromagnetism of Maxwell's equations and applies to most superpositions of laser pulses.  We showed how the invariant mass density and the rapidity of the moving rest frame are controlled.  For the example of two colliding high-intensity square laser pulses, we showed how to create collimated high energy electron-positron bunches that are directed away from both laser beam lines and may be directed independently after creation.  The resulting bunches are planar in the plane normal to the electric field and Gaussian in momentum.  

We have discussed how the total energy of the unstable field configuration can be smoothly allotted between the energy per particle and total number of particles in the bunch.  In the present model, the particle number in the resultant bunch depends most on the absolute magnitude of the fields attained in the collision volume.  Techniques such as optimizing pulse shape~\cite{DiPiazza04} and frequency structure~\cite{Dunne:2009gi,Dumlu:2010ua}  can be applied to enhance the yield. 

Particles are produced in their rest frame at a very low energy, and the high laboratory energy is achieved by superposing nearly collinear, pulsed electromagnetic fields so as to have the rest frame moving at high rapidity relative to the lab.  
No acceleration process is needed to bring the charged particles to high energy. In our proposed beam geometry the laser fields may be made very strong without endangering the stability of the experiment or its apparatus by near-critical laser pulses.  The vacuum-decay products in laser pulse collisions are readily distinguishable from particle-creation backgrounds such as cascades~\cite{Fedotov:2010ja} and perturbative processes thus enhancing discovery potential.  The concept we have described may permit future development of a source of particles with ultra high energy.  

\vskip 0.2cm
\section*{Acknowledgments} 
This work was supported by a grant from the 
U.S. Department of Energy, DE-FG02-04ER41318.

\end{document}